\documentclass[aps,prb,twocolumn,showpacs,showkeys]{revtex4}

\usepackage{bm}% bold math
\usepackage{amsmath}%ams
\usepackage{t1enc}              % polish letters
\usepackage[cp1250]{inputenc}   % polish keyboard
\usepackage[english]{babel}

\usepackage{graphicx}

\begin{document}

%\preprint{}

\title{Surface states in zigzag and armchair graphene nanoribbons }


\author{Jarosław Kłos}
\email{klos@amu.edu.pl}
%\thanks{}
\affiliation{Surface Physics Division, Faculty of Physics, Adam
Mickiewicz University,  ul. Umultowska 85, 61-614 Poznań, Poland\\
Department of Science and Technology, Linköping University
 601 74, Norrköping, Sweden
}
%\thanks{}
%\affiliation{ }


\date{\today}

\begin{abstract}
This paper presents electronic spectra of zigzag and armchair graphene nanoribbons calculated within the tight-binding model for $\pi$-electrons. Zigzag and armchair nanoribbons of different edge geometries are considered, with surface perturbation taken into account. The properties of surface states are discussed on the basis of their classification into Tamm states and Shockley states. In armchair nanoribbons surface states are shown to close the energy gap at the Dirac point for certain edge geometries.
\end{abstract}

%Electron states at surfaces and interfaces, 
%Electronic structure of nanoscale materials: clusters, nanoparticles,nanotubes, and nanocrystal
%Quantum wires

\maketitle

\section{Introduction}
Graphite-based materials are the subject of intensive research for both fundamental and practical reasons. Fullerenes\cite{fuller}, carbon nanotubes\cite{nanotubes3,nanotubes2} and graphene nanoribbons\cite{ribbon-mag,graphen-rev} have unusual electronic properties, which allow the design and fabrication of nanoelectronic systems with parameters unavailable to conventional electronics in this size range\cite{graphene-electonics}. 

Graphene, a single layer of graphite, is composed of carbon atoms arranged in a two-dimensional honeycomb lattice. Covalent $\sigma$ bonds between hybridized $sp_{2}$ orbitals form the skeleton of a graphene sheet. Electronic transport in graphene is, in principle, based on electrons from $p_{z}$ orbitals forming delocalized $\pi$ bonds. The tight-binding approximation (TBA) commonly used for a rough description of transport properties of graphite-based materials\cite{TBA-graphene,TBA-kp} only takes into account the nearest-neighbor hopping of $\pi$ electrons.

The unit cell of the honeycomb lattice comprises two lattice sites. Consequently, two energy bands of graphene are obtained in the tight-binding model with hopping limited to nearest neighbors. The bands touch at six $K$ points in the Brillouin zone. Exactly between the bands, the Fermi energy passes through the $K$ points. The linear character of the dispersion relation near the $K$ points results in electrons behaving as massless fermions\cite{massless-ferm} described by the Dirac equation\cite{dirac}; hence the $K$ points are referred to as Dirac points. As a two-dimensional system with a specific topology of the band structure, graphene is characterized by the occurrence of non-zero Berry phases of the electronic wave function\cite{berry-ph}. This brings about both the integer and the fractional quantum Hall effect, predicted theoretically\cite{QHE-SdH,QHE2} and verified experimentally\cite{QHE-SdH-exp,QHE-exp}.

Carbon nanotubes and graphene nanoribbons are quasi-1D systems formed by rolling or confining, respectively, a graphene sheet along the direction defined by the so-called chiral vector $v_{c}$. The assumed boundary conditions imply the quantization of the wave vector component $k_{c}$ describing the propagation of an electron wave in graphene in the direction of $v_{c}$. The corresponding dispersion relation $E_{n}(k_{\parallel})$ for the propagation along the nanotube/nanoribbon comprises a number of branches resulting from the quantization of $k_{c}$\cite{conduct-quant}. This implies quantized conductance in carbon nanotubes and graphene nanoribbons. At low temperatures electronic transport in quasi-1D graphene structures presents spin-related effects, such as the Coulomb blockade\cite{coul-block-ribb,coul-block-dot} or the Kondo effect\cite{kondo}, due to the electron-electron interaction.

In carbon nanotubes all the dispersion branches lie within the projection of the 2D graphene dispersion relation on the direction of $k_{c}$, as implied by the periodic boundary conditions and the lack of surface\cite{nanotubes2}. In graphene nanoribbons, due to their finite size in the direction of $v_{c}$, the quantized wave vector component $k_{c}$ can take on complex values. The corresponding states localize at the nanoribbon surface(edge)\cite{st-edge,st-edge2,st-edge-med} and their dispersion branches $E_{n}(k_{\parallel})$ lie beyond the projection of the graphene dispersion relation on the direction of $k_{c}$.

The orientation of $k_{c}$ in the Brillouin zone also determines the position of the Dirac points in the dispersion relation $E_{n}(k_{\parallel})$. For $k_{c}$ oriented along the $\Gamma-M$ direction Dirac points occur at $k_{\parallel}=\pm\tfrac{2}{3}\pi$ . Rotating $k_{c}$ to the $\Gamma-K$ direction results in the Dirac points shifting to $k_{\parallel}=0$.

The transport properties are determined by the electronic structure near the Fermi energy. As in graphene the Fermi energy passes through the Dirac points, the orientation of $k_{c}$ with respect to the Dirac points, and the circumference/length of the nanotube/nanoribbon are decisive for its metallic or semiconducting character. By suitable adjustment of these parameters $k_{c}$ can be quantized in a manner which implies some dispersion branches passing through the Dirac points. This results in the energy gap being closed by branches of bulk states.

In nanoribbons the energy gap can be closed by another mechanism as well. Metallic character of the systems (by assuming simple TBA model) can be due to the occurrence of surface states\cite{st-edge2} between bands of bulk states. In zigzag graphene nanoribbons (with $k_{c}$ oriented in the $\Gamma-M$ direction) strongly localized surface states occur near the Fermi energy. The role of these states is especially important in thin nanoribbons, with a high density of surface states in a relatively wide gap between the upper-band bulk states and the lower-band ones. No surface states have been shown to occur in graphene nanoribbons of pure armchair edge structure ($k_{c}$ along the $\Gamma-K$ direction) in the tight-binding approach\cite{st-edge}. For $k_{c}$ orientations between the $\Gamma-K$ and $\Gamma-M$ directions, dimers of carbon atoms (characteristic of the armchair orientation) as well as single atoms (typical of the zigzag orientation) occur at the nanoribbon edges\cite{st-edge}. Surface states have been shown to localize at surface atoms typical of the zigzag edge and their density to have a peak at the Fermi energy. Interestingly, the nearly flat dispersion branches $E_{n}(k_{\parallel})$ in the vicinity of $E=0$ and the consequent high density of states imply spin-polarized edges when Hubbard repulsion is taken into account\cite{st-edge2,TBA-hubb}. Spin-polarization results in half-metallic properties of zigzag nanoribbon\cite{half-met}. It is worth to note that the result of the electron-electron interaction is the opening a small gap at Fermi level both for zigzag and 'metallic' armchair nanoribons\cite{gap-w}. This processes were not taken into account in the presented paper.

Two factors determine the occurrence of surface states in the system considered in the model approach: broken translational symmetry and surface perturbation. These two factors provide the basis for the distinction, used by some authors \cite{artmann,zak,klos}, of two categories of surface states, referred to as Shockley\cite{shockley} states and Tamm\cite{tamm} states. Shockley states are surface states predicted to occur at a non-reconstructed surface conserving the chemical composition of the bulk. The conditions of existence of Shockley states at a non-perturbed surface can be formulated on the basis of the symmetry of the system. The surface perturbation, when taken into account, tends to impair the localization of surface states of this type. In contrast to Shockley states, Tamm states necessitate a surface perturbation to occur. Real surfaces are always perturbed due to reconstruction and chemisorption processes. However, in spite of the concurrent occurrence of broken translational symmetry and surface perturbation, the discussed classification can be of use for studying the conditions of existence of surface states in the system.

Many studies of graphene nanoribbons in the tight-binding approach assume no surface reconstruction and no significant effect of hydrogen passivation of the surface on the energy of surface carbon atoms. The main purpose of the passivation, performed in experimental studies, is to saturate the $sp_{2}$ bonds of surface carbon atoms. As a result, zigzag edges are composed of alternately arranged surface and bulk carbon atoms (having two and three neighbors, respectively), while armchair edges consist of alternately placed dimers of surface atoms and dimers of bulk atoms. 

A free radical in the form of an extra hydrogen atom or a methyl group can be added or at the nanoribbon surface. This addition results in modified hybridization of surface carbon atom orbitals, which become $sp_{3}$ hybrids\cite{surf-H-bound-dys,surf-H}. There is also possible to consider saturation of some dangling carbon bonds by methylene groups\cite{CH2-fp} (not by hydrogen). This delivers additional $\pi$-electrons coming from each $-CH_{2}$ group. Processes in which the chemical composition of the surface is modified can be considered, too. It can by done by substituting boron atoms or nitrogen atoms in place of carbon atoms\cite{NOB-DFT} or by passivation of surface by another radicals: $-F$, $-0$, $-OH$\cite{H-OH-F-fp}. These effects can be taken into account in the tight-binding model by modifying the energy of a surface carbon atom\cite{st-edge-med} and its neighbor hooping, or, more radically, by excluding a carbon atom from the surface (or adding an extra one). This will modify the geometry of the nanoribbon skeleton in which electrons propagate. In the resulting model  each surface atom will have only neighbor with one bulk carbon atom.

In this paper surface states localized at graphene nanoribbon edges are determined in the tight-binding approximation with nearest-neighbor hopping. Different model geometries of armchair and zigzag nanoribbons are considered, with surface carbon atoms having two bulk neighbors or a single bulk neighbor. The surface perturbation is modeled by a modification of the surface atom energy. The effect of energy gap closing by bulk states or surface states in armchair and zigzag nanoribbons and its dependence on the edge geometry are discussed as well. 

The ensuing part of this paper is organized as follows: Section II discusses the considered graphene nanoribbon structures. Section III presents the method employed in the calculations, the results of which are presented and discussed in Section IV. The study is summed up in the closing Section V.

\section{Model structures}
\begin{figure}
\centering
\includegraphics[width=3.45in]{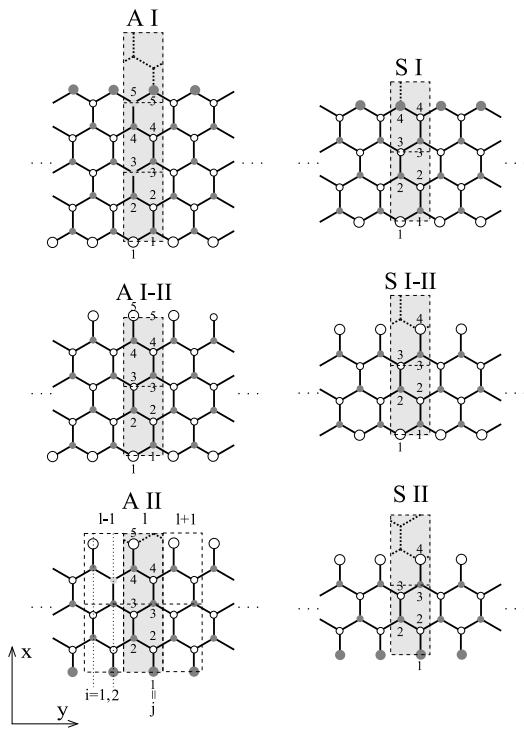}
\caption{\label{fig:f1}Structure of 
$\pi$-electron skeleton of zigzag graphene nanoribbon. Circles and lines represent lattice sites and hoppings, respectively; large circles represent surface (edge) sites. Each surface site has two neighbors (in edge configuration $I$) or a single neighbor (in edge configuration $II$). Structures $AI$ and $AII$ have a center of symmetry, while structures $SI$ and $SII$ have a symmetry axis. Structures $AI-II$ ($AI$ combined with $AII$) and $SI-II$ ($SI$ combined with $SII$) have edges of both types. Dashed lines are limits of rectangular unit cells of graphene sheet (periodic in $x$ and $y$ directions); gray area represents a cell of graphene nanoribbon (periodic in y direction). The nanoribbon cell comprises two slices of lattice sites. Surface perturbation is introduced by modifying the energy (potential) of surface sites.}
\end{figure}

\begin{figure}
\centering
\includegraphics[width=3.45in]{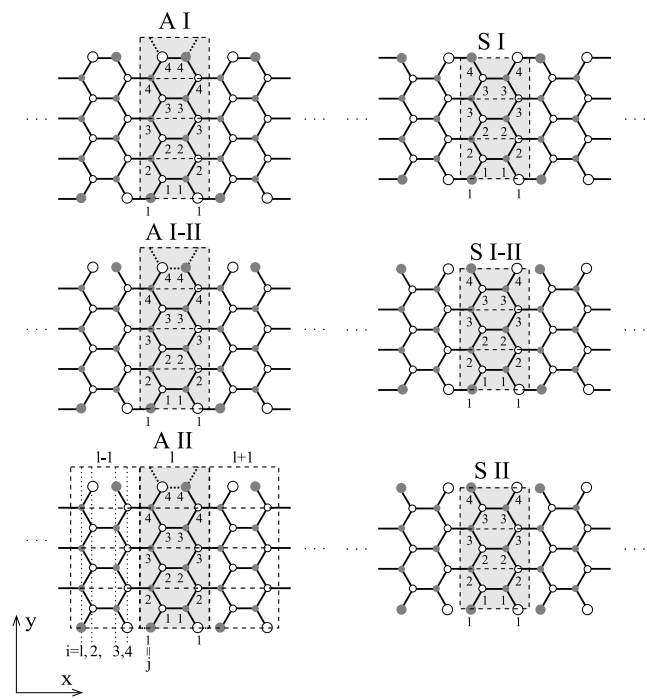}
\caption{\label{fig:f2}Structure of $\pi$-electron skeleton of armchair graphene nanoribbon. Symbols and classification rules are as for the zigzag structure (Fig. \ref{fig:f1}). Note the nanoribbon cell (periodic in $x$ direction) comprises four slices of lattice sites.}
\end{figure}

The structures depicted in Figs.\ref{fig:f1} and \ref{fig:f2} represent the skeleton of graphene nanoribbons with  $\pi$-electron delocalization. Figures \ref{fig:f1} and \ref{fig:f2} provide a scheme for the description of the electronic properties of the system in the tight-binding model; circles and lines represent lattice sites and hoppings between them, respectively.

Our interest is limited to structures without surface reconstruction. The geometry of a reconstructed surface can differ substantially from that of the nanoribbon bulk and practically represent a different phase\cite{recons-fp}. Quite challenging, a systematic description of such systems can be the subject of further studies.

The nanoribbons shown in Figs. \ref{fig:f1} and \ref{fig:f2} are cut from a graphene sheet in two particular directions, resulting in two types of edge termination, referred to as zigzag and armchair. In graphene nanoribbons with other orientations of the chiral vector the edge structure has many steps and terraces\cite{st-edge-STM,st-edge-med-DFT} and can be regarded as a mixture of the zigzag structure and the armchair one.

Let us consider two model edge configurations of the nanoribbon skeleton, labeled $I$ and $II$ in Figs. \ref{fig:f1} and \ref{fig:f2}. In the type $I$ configuration, each surface (edge) site has two bulk neighbors, while in the type $II$ configuration each surface site has only one bulk neighbor. The type $I$ configuration, with surface carbon atoms $sp_{2}$-hybridized as a result of hydrogen saturation of the broken bond, is regarded in the literature as that of a non-reconstructed pure surface of graphene nanoribbon. However, a free radical added to a hydrogen-passivated surface atom makes it $sp_{3}$-hybridized and excluded from the skeleton of sites participating in transport of  $\pi$-electrons. Regardless of the physical mechanism, discussed in the previous section, we can presume that surface sites with a single bulk neighbor should be taken into account in the tight-binding model. Configuration $II$ represents a limiting case. In real systems only a part of surface atoms have a single neighbor, due to reduced radical-carbon bond dissociation energy or steric restrictions, which increase with coverage of the surface by radicals/cite{surf-H-bound-dys}.

Depending of the nanoribbon width, nanoribbons with both edges of the same configuration ($I$ or $II$) can have a symmetry axis (structures $SI$ and $SII$) or a center of symmetry (structures $AI$ and $AII$). Systems with 'mixed' edges (structures $AI-II$ and $SI-II$) are easily seen to have neither a symmetry axis along the nanoribbon nor a center of symmetry.

\begin{figure}
\centering
\includegraphics[width=2.5in]{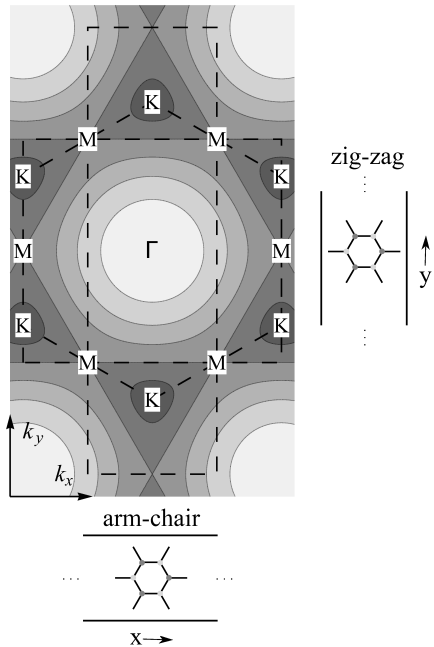}
\caption{\label{fig:f3}Two-dimensional dispersion relation of graphene for upper/lower band; dark areas correspond to energy values close to zero. The hexagon with vertices at $K$ points delimits the first Brillouin zone. Inside, the rectangle with vertices at $M$ points represents the folded Brillouin zone corresponding to the rectangular unit cell of the original lattice. The elongated rectangles are areas of projection of dispersion relation on $\Gamma-K$ and $\Gamma-M$ directions. Bottom and right, graphene lattice against edges of armchair and zigzag nanoribbons, respectively.}
\end{figure}

A hexagonal unit cell comprising two atoms can be defined in an infinite graphene sheet. Repeated unit translations of the unit cell atoms produce two sublattices, distinguished by empty and filled circles in Figs. \ref{fig:f1} and \ref{fig:f2}. Note that a different lattice cell is more convenient in the case of graphene nanoribbon. A rectangular cell comprising four atoms reproduces the structure of the nanoribbon when translated along and perpendicularly to its axis. The area of this nanoribbon cell is twice as large as that of the hexagonal unit cell of graphene. Consequently, the corresponding rectangular Brillouin zone (with vertices in $M$ points) is formed by folding the hexagonal Brillouin zone (with vertices at $K$ points), see Fig. \ref{fig:f3}. The area covered by the bulk dispersion branches $E_{n}(k_{\parallel})$ of an armchair or zigzag nanoribbon is determined by a simple projection of the 2D dispersion relation of graphene on the $\Gamma-K$ direction or the $\Gamma-M$ direction, respectively, from the areas delimited by elongated rectangular frames in Fig. \ref{fig:f3}. This operation is equivalent to a projection of the folded Brillouin zone. 

In our discussion of the geometry of graphene nanoribbons we have not taken into account surface perturbation so far. We have only allowed for different orientations (armchair or zigzag) of the nanoribbon against the graphene lattice and different positions of edges in the unit cell (configurations $I$ and $II$). Graphene nanoribbons are often regarded as unrolled carbon nanotubes. It is worthy of notice that among the systems under consideration only zigzag structures $SI$ and $AII$ and armchair structure $AI$ can be rolled to form a carbon nanotube. These structures have an integer number of rectangular unit cells (delimited by dashed line in Figs. \ref{fig:f1} and \ref{fig:f2}) across the nanoribbon. Below we shall attempt to determine which of the model structures under consideration permit the existence of Shockley states, or surface states which do not require surface perturbation.

In the considered model the surface perturbation will be introduced by modifying the energy (potential) of surface atoms in the  $\pi$-electron skeleton. In Figs. \ref{fig:f1} and \ref{fig:f2} this is visualized by enlarged circles representing surface sites. We are going to determine the effect of surface perturbation defined in this manner on the existence of Shockley states, as well as the conditions in which Tamm states appear.

\section{Method}
The method presented below allows to determine wave vectors of modes of known energy, propagating in the nanoribbon along its axis. First developed for the determination of band structure of lattices of mesoscopic quantum dots\cite{green-rec}, this technique was subsequently adapted to the calculation of graphene nanoribbon spectra\cite{trans-mag-barr}.

The following symbols and units will be used below for clarity of presentation: The wave-vector components $k_{x}$ and $k_{y}$ will be expressed in units of $\tfrac{1}{3a}$ and $\tfrac{1}{\sqrt{3}a}$, respectively, where $3a$ and $\sqrt{3}a$ are dimensions of the rectangular unit cell of the graphene lattice and $a$ is the bond length. The obtained magnitudes of the dimensionless components $k_{x}$ and $k_{y}$ in the folded Brillouin zone range from $-\pi$ to $\pi$. The electron energy $X$ is expressed in units of the hopping integral $t$. The assumed zero energy level is the energy (potential) in lattice sites $\epsilon$
\begin{equation}
X=\frac{E-\epsilon}{t}.
\end{equation}

Considering the translational symmetry along the nanoribbon, let us define a  cell comprising $M=2$ or $M=4$ slices of lattice sites in the zigzag structure and in the armchair structure, respectively (see Figs. \ref{fig:f1} and \ref{fig:f2}). The tight-binding Hamiltonian can be expressed in the form of a block tridiagonal supermatrix in which the diagonal submatrices (blocks) represent Hamiltonians of isolated cells:
\begin{equation}
\bm{H}_{l,l}=\bm{H}^{cell}=\left(
\begin{array}{ccccc}
H_{1}&&&\cdots&0\\
\ddots&\ddots&\ddots&&\vdots\\
&U_{i,i-1}&H_{i}&U_{i,i+1}&\\
\vdots&&\ddots&\ddots&\ddots\\
0&\cdots&&&H_{M}
\end{array}
\right)
\end{equation}
and the superdiagonal $\bm{H}_{l,l+1}=\bm{U}$ and subdiagonal  $\bm{H}_{l+1,l}=\bm{U}^{\dagger}$ blocks define the intercell hopping: 
\begin{equation}
\bm{U}=\left(
\begin{array}{ccccc}
0&\cdots &&0&U_{1,M}\\
\vdots&\ddots&&&0\\
&&0&&\\
0&&&\ddots&\vdots\\
U_{M,1}&0&&\cdots&0
\end{array}
\right).
\end{equation}
Element $H_{i}$ is the Hamiltonian matrix of the $i$-th isolated slice of lattice sites, and $U_{i,i\pm 1}$ is the hopping matrix for adjacent slices. Note matrices $U_{i,i\pm i}$ are generally rectangular, due to the possibly different number $N_{j}$ of lattice sites in different slices. In the adopted units of energy all the diagonal elements $H_{i}^{j,j}$ are zero, with the exception of $H_{i}^{1,1}$ and $H_{i}^{Nj,Nj}$, which take on values:
\begin{equation}
Z=\frac{\epsilon_{s}-\epsilon}{t},
\end{equation}
if the extreme sites in the slice $i$ are surface sites ($\epsilon_{s}$ denotes energy of surface site. The nonzero elements of $H_{i}$ and $U_{i,i\pm 1}$ indicate intersite hopping within the slice or between adjacent slices, respectively. 

Let us divide the state space into two subspaces, one referring to a reference cell and the other to the rest of the system:
\begin{equation}
\left|\Psi\right\rangle=\left|\Psi_{cell}\right\rangle\oplus\left|\Psi_{out}\right\rangle.\label{eq:e5}
\end{equation}
The Hamiltonian of the whole system can be formally written as:
\begin{equation}
\bm{H}=\bm{H}^{cell}+\bm{H}^{out}+\tilde{\bm{U}},\label{eq:e6}
\end{equation}
where $\tilde{\bm{U}}$ defines hopping between the reference cell and the rest of the system, defined by the Hamiltonian $\bm{H}^{out}$. All matrices are written in a whole space of $\left|\Psi\right\rangle$. 

By using the standard definition of a Green's function:
\begin{equation}
\bm{G}^{cell}=(X\hat{\bm{1}}-\bm{H}^{cell})^{-1}
\end{equation}
combined with (\ref{eq:e5}) and (\ref{eq:e6}) and Sch\"{o}dinger equation:
\begin{equation}
\bm{H}\left|\Psi\right\rangle=X\left|\Psi\right\rangle,
\end{equation}
the following relation can be derived: 
\begin{equation}
\left|\Psi_{cell}\right\rangle=\bm{G}^{cell}\tilde{\bm{U}}\left|\Psi_{out}\right\rangle.\label{eq:e9}
\end{equation}
$\bm{G}^{cell}$ can be determined by the recursive Green's function technique\cite{green-rec2}. 

The relation (\ref{eq:e9}) allows to interrelate the functions at the interfaces between the reference cell ($l$-th) and the rest of the system\cite{trans-mag-barr}:
\begin{eqnarray}
\Psi^{l}_{1}&=&G^{cell}_{1,1}U_{1,M}\Psi^{l-1}_{M}+G^{cell}_{1,M}U_{M,1}\Psi^{l+1}_{1},\nonumber\\
\Psi^{l}_{M}&=&G^{cell}_{M,1}U_{1,M}\Psi^{l-1}_{M}+G^{cell}_{M,M}U_{M,1}\Psi^{l+1}_{1},\label{eq:e10}
\end{eqnarray}
where columns of the matrix $\Psi^{l}_{i}$ represent wave functions for the $i$-th slice in the $l$-th cell. The system of equations (\ref{eq:e10}) can be put in the matrix form:
\begin{equation}
T_{1}\left(
\begin{array}{c}
\Psi^{l+1}_{1}\\
\Psi^{l}_{M}
\end{array}
\right)=
T_{2}\left(
\begin{array}{c}
\Psi^{l}_{1}\\
\Psi^{l-1}_{M}
\end{array}
\right),
\end{equation}
where
\begin{eqnarray}
T_{1}&=&\left(
\begin{array}{cc}
-G^{cell}_{1,M}U_{M,1}&\hat{0}\\
-G^{cell}_{M,M}U_{M,1}&\hat{1}
\end{array}
\right),\nonumber
\\
T_{2}&=&\left(
\begin{array}{cc}
-\hat{1}&G^{cell}_{1,1}U_{1,M}\\
\hat{0}&G^{cell}_{M,1}U_{1,M}
\end{array}
\right).
\end{eqnarray}
By applying Bloch's theorem:
\begin{equation}
\Psi^{l+1}_{i}=e^{\bm{i} k_{\parallel}}\Psi^{l}_{i}
\end{equation}
we obtain a generalized eigenvalue problem, which allows the determination of wave vectors $k_{\parallel}$ of nanoribbon modes of known energy $X$:
\begin{equation}
T_{2}\left(
\begin{array}{c}
\Psi^{l}_{1}\\
\Psi^{l-1}_{M}
\end{array}
\right)=e^{\bm{i} k_{\parallel}}T_{1}\left(
\begin{array}{c}
\Psi^{l}_{1}\\
\Psi^{l-1}_{M}
\end{array}
\right)\label{eq:e14}
\end{equation}
The solutions of the eigenvalue problem (\ref{eq:e14}) include evanescent modes, characterized by complex values of $k_{\parallel}$. As the system is infinite along the nanoribbon axis, these solutions must be discarded as non-physical. Note the eigensolver finds solutions corresponding to modes propagating (or evanescent) in each of the opposite directions, and thus with opposite signs of $k_{\parallel}$.

To determine the localization of states in the direction perpendicular to the nanoribbon axis we must check whether the assumed state energy $X$ for the determined value of $k_{\parallel}$ is within the energy gap of the projection of the graphene dispersion relation. The following conditions of existence of surface states are obtained for the zigzag orientation:
\begin{equation}
\left|X\right|>1+2\cos(\tfrac{1}{2}k_{y})
\end{equation}
for states below the lower band and above the upper one, and:
\begin{eqnarray}
\left|X\right|<1-2\cos(\tfrac{1}{2}k_{y})&\rm{for}&\left|k_{y}\right|>\tfrac{2}{3}\pi,\nonumber\\
\left|X\right|<2\cos(\tfrac{1}{2}k_{y})-1&\rm{for}&\left|k_{y}\right|<\tfrac{2}{3}\pi
\end{eqnarray}
between the bands. For the armchair orientation the conditions become:
\begin{equation}
\left|X\right|>\sqrt{5+4\cos(\tfrac{1}{2}k_{x})}
\end{equation}
below the lower band and above the upper one, and:
\begin{eqnarray}
\left|X\right|<\sqrt{3-2\sin(\tfrac{1}{6}\pi-\tfrac{2}{3}k_{x})-4\sin(\tfrac{1}{3}k_{x}+\tfrac{1}{6}\pi)}\nonumber\\
{\rm for}\;k_{x}>0,\nonumber\\
\left|X\right|<\sqrt{3-2\sin(\tfrac{1}{6}\pi+\tfrac{2}{3}k_{x})+4\sin(\tfrac{1}{3}k_{x} - \tfrac{1}{6}\pi)}\nonumber\\
{\rm for}\;k_{x}<0
\end{eqnarray}
between the bands.

\section{Results}
\begin{table}[b]
\caption{\label{tab:t1}Number $N$ of dispersion branches in spectrum and number $M$ of modes at fixed energy value in nanoribbons of different width $n$, orientation and edge geometry.}
\begin{ruledtabular}
\begin{tabular}{c|cccccc}
&\multicolumn{6}{c}{zig-zag} \\
structure &AI&AI-II&AII&SI&SI-II&SII \\
\hline
N&2n&2n-1&2n-2&2n&2n-1&2n-2\\
M&2n&2n-2&2n-4&2n&2n-2&2n-4\\
\hline 
&\multicolumn{6}{c}{arm-chair} \\
structure &AI&AI-II&AII&SI&SI-II&SII \\
\hline
N&4n&4n&4n&4n-2&4n-2&4n-2\\
M&2n&2n-2&2n-2&2n-2&2n-2&2n-4\\
\end{tabular}
\end{ruledtabular}
\end{table}

\begin{figure}
\centering
\includegraphics[width=3.5in]{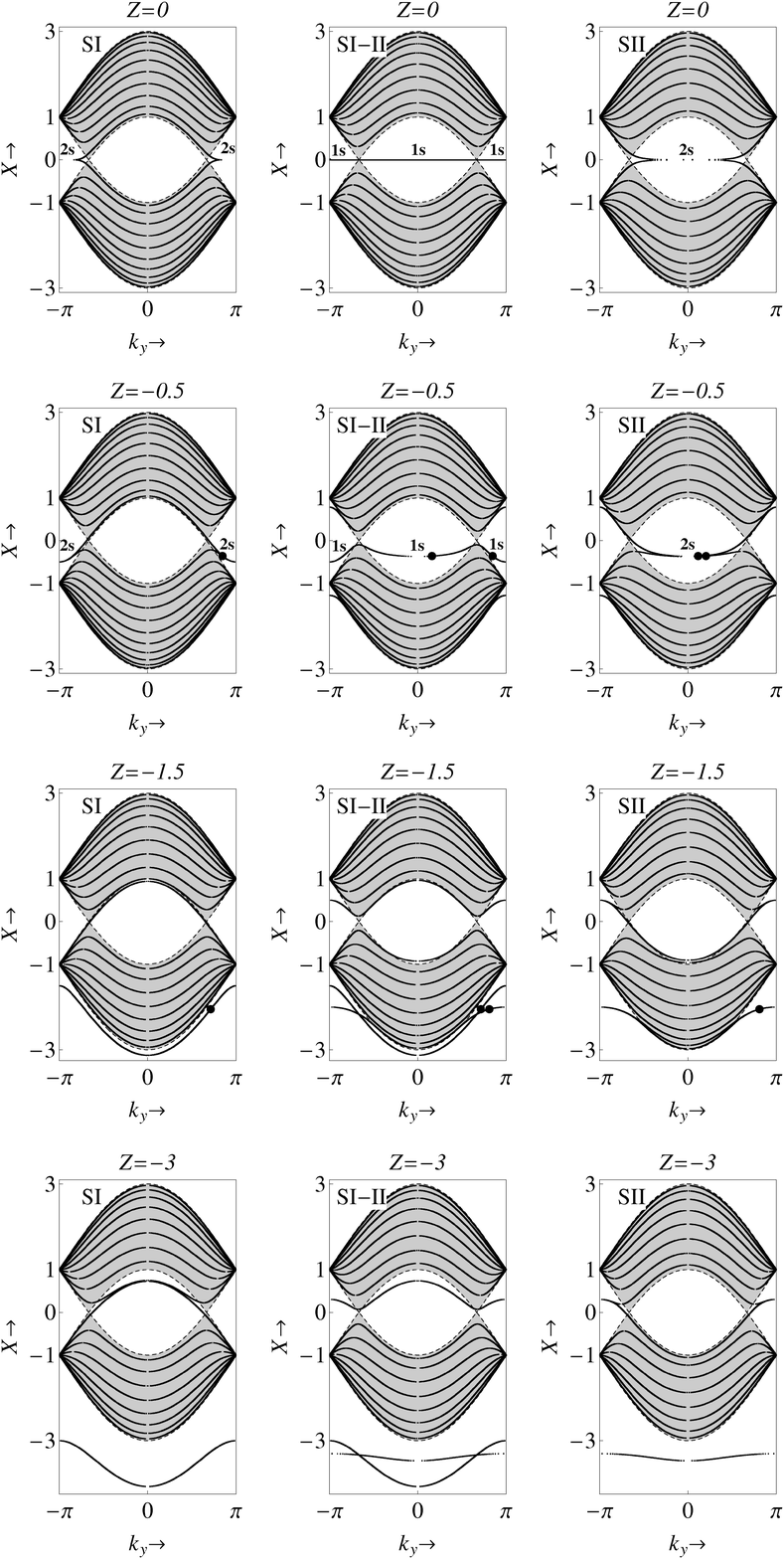}
\caption{\label{fig:f4}
Energy spectra of zigzag graphene nanoribbons of width $n=10$. Plots obtained for geometries $SI$, $SI-II$ and $SII$ are shown in respective columns. Spectra in successive rows correspond to growing absolute value of surface perturbation: $Z=0$, $-0.5$, $-1.5$ and $-3$. Thick lines, composed of calculation points, represent dispersion branches corresponding to each mode; spacing between points in flattened branches is due to finite energy step in the calculation procedure. Grey area represents projection of the 2D graphene dispersion relation on the direction of $k_{x}$ (cf. Fig. \ref{fig:f3}). Dots in surface state dispersion branches correspond to $k_{y}$ and $X$ values assumed in the wave function plots in Fig. \ref{fig:f5}. Labels: $1s$ and $2s$ mark one or two branches of Shockley states respectively.}
\end{figure}

\begin{figure}
\centering
\includegraphics[width=3.5in]{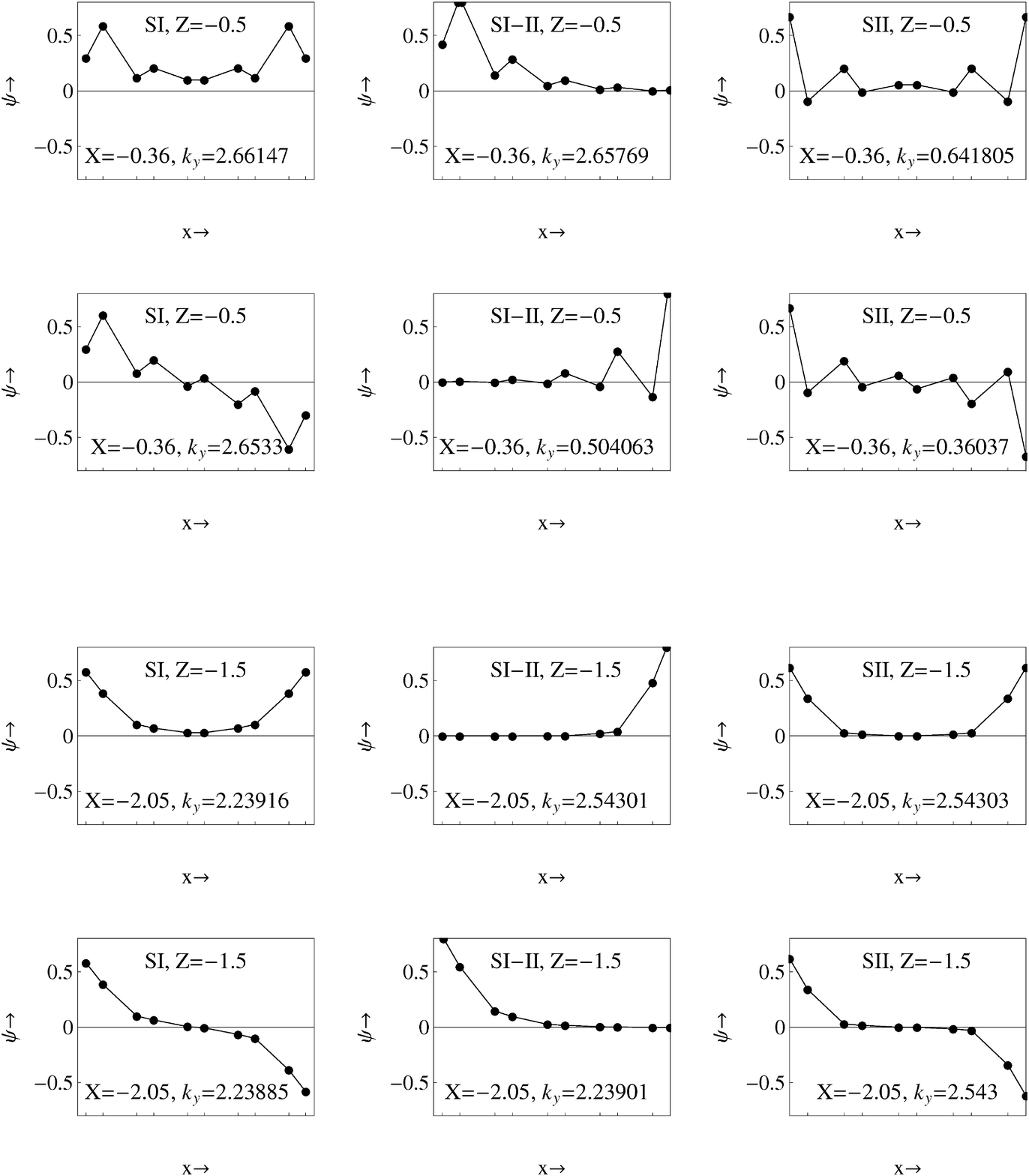}
\caption{\label{fig:f5}
Profiles of surface state wave functions plotted across a zigzag nanoribbon for the first slices of sites in the nanoribbon cell (cf. Fig. \ref{fig:f1}). The states are indicated in dispersion spectra plotted in Fig. \ref{fig:f4} by black dots.} 
\end{figure}

Before proceeding to the presentation of results, let us discuss the effect of the nanoribbon width on the number of dispersion branches and the number of solutions of the generalized eigenvalue problem. As the measure of nanoribbon width let us assume the maximum number $n$ of sites in a slice, allowing for possible slice shift (e.g. for structure $AII$ in Fig. \ref{fig:f1} we assume $n=5$). The number of dispersion branches and the number of solutions of the eigenvalue problem for different nanoribbon widths are specified in Table \ref{tab:t1}. The number of dispersion branches is easily seen to be equal to that of sites in the nanoribbon cell (gray area in Figs. \ref{fig:f1} and \ref{fig:f2}). The number of solutions of the general eigenvalue problem determines the maximum number of modes propagating in the nanoribbon at a fixed value of energy $X$: the number of propagating modes is the total number of solutions of the eigenvalue problem minus the number of non-physical solutions, corresponding to evanescent modes.

Figure \ref{fig:f4} shows spectra obtained for zigzag nanoribbons. Only spectra of structures $SI$, $SI-II$ and $SII$ are depicted, those of structures $AI$, $AI-II$ and $AII$ presenting no significant differences. The spectra were calculated for nanoribbon width $n=10$, which is reflected in the number of dispersion branches: $20$, $19$ and $18$ in structures $SI$, $SI-II$ and $SII$, respectively. Shown in rows corresponding to growing surface perturbation, the spectra prove symmetric with respect to the sign of surface perturbation: its reversal, $Z\rightarrow-Z$, results in a spectrum reflected with respect to $X=0$.
 
The gray area in Fig. \ref{fig:f4} represents the projection of the graphene dispersion relation on the direction of $k_{x}$. Branches beyond this area correspond to surface states. Shockley surface states are seen to occur at both types ($I$ and $II$) of non-perturbed surface ($Z=0$). For the type I edge configuration Shockley states appear in the band gap for $\left|k_{y}\right|<\tfrac{2}{3}\pi$. If the gap is wide, the value of the imaginary component of the wave vector is high, which implies strongly localized states. This reduces the interaction between opposite edges of the nanoribbon and causes a gradual degeneracy of even and odd states in the symmetric structures $SI$ and $SII$ (cf. Fig. \ref{fig:f5}). For $k_{y}=\pm\pi$ Shockley states at the type $I$ surface are localized at a surface site and isolated from the rest of the system, hence their energy $X=Z$. 

\begin{figure}
\centering
\includegraphics[width=3.5in]{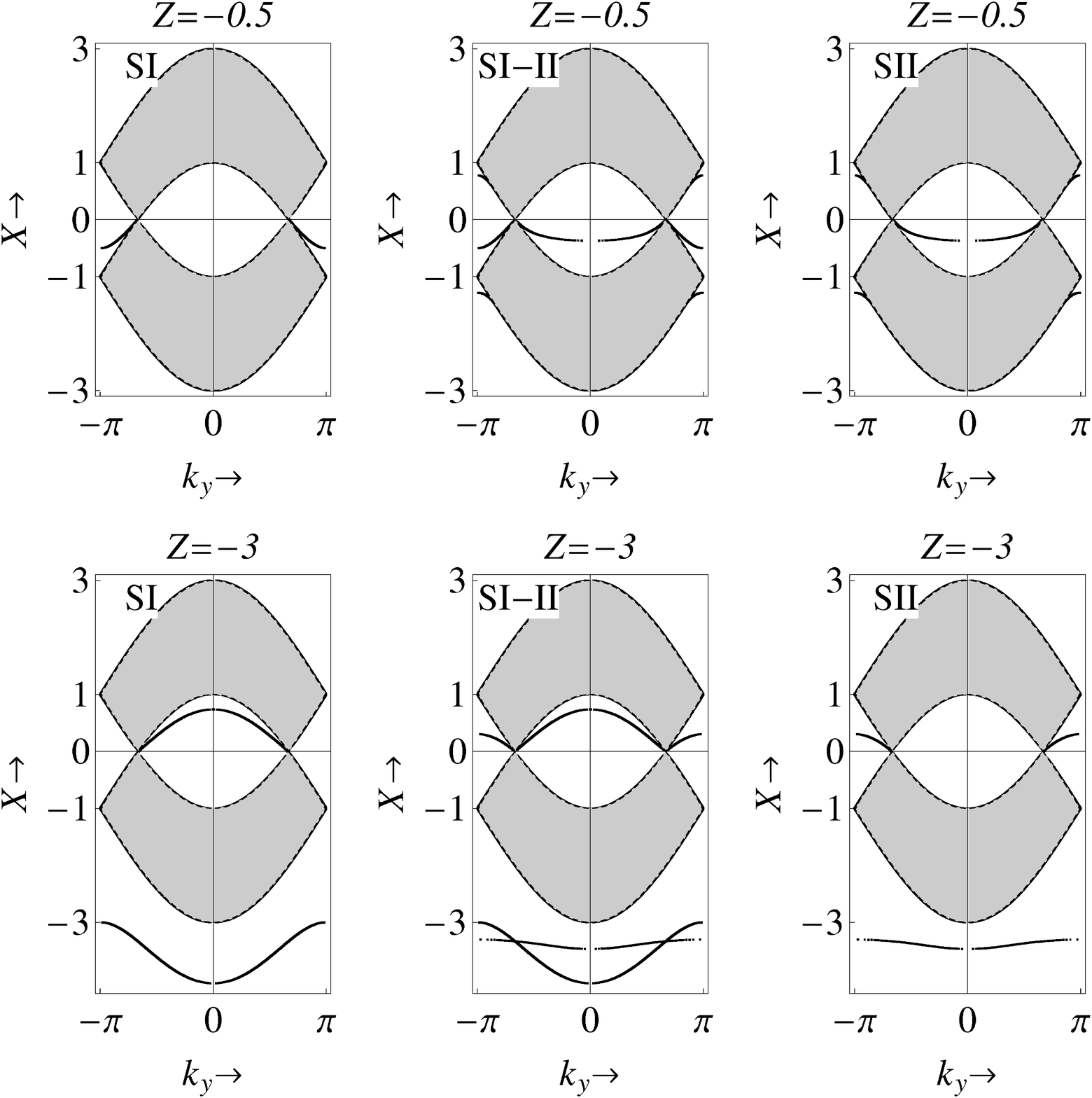}
\caption{\label{fig:f6}Dispersion spectra of surface states in zigzag nanoribbons of width $n=50$ and edge geometries $SI$, $SI-II$ and $SII$ with surface perturbation $Z=-0.5$ and $-3$.}
\end{figure}

\begin{figure}
\centering
\includegraphics[width=3.5in]{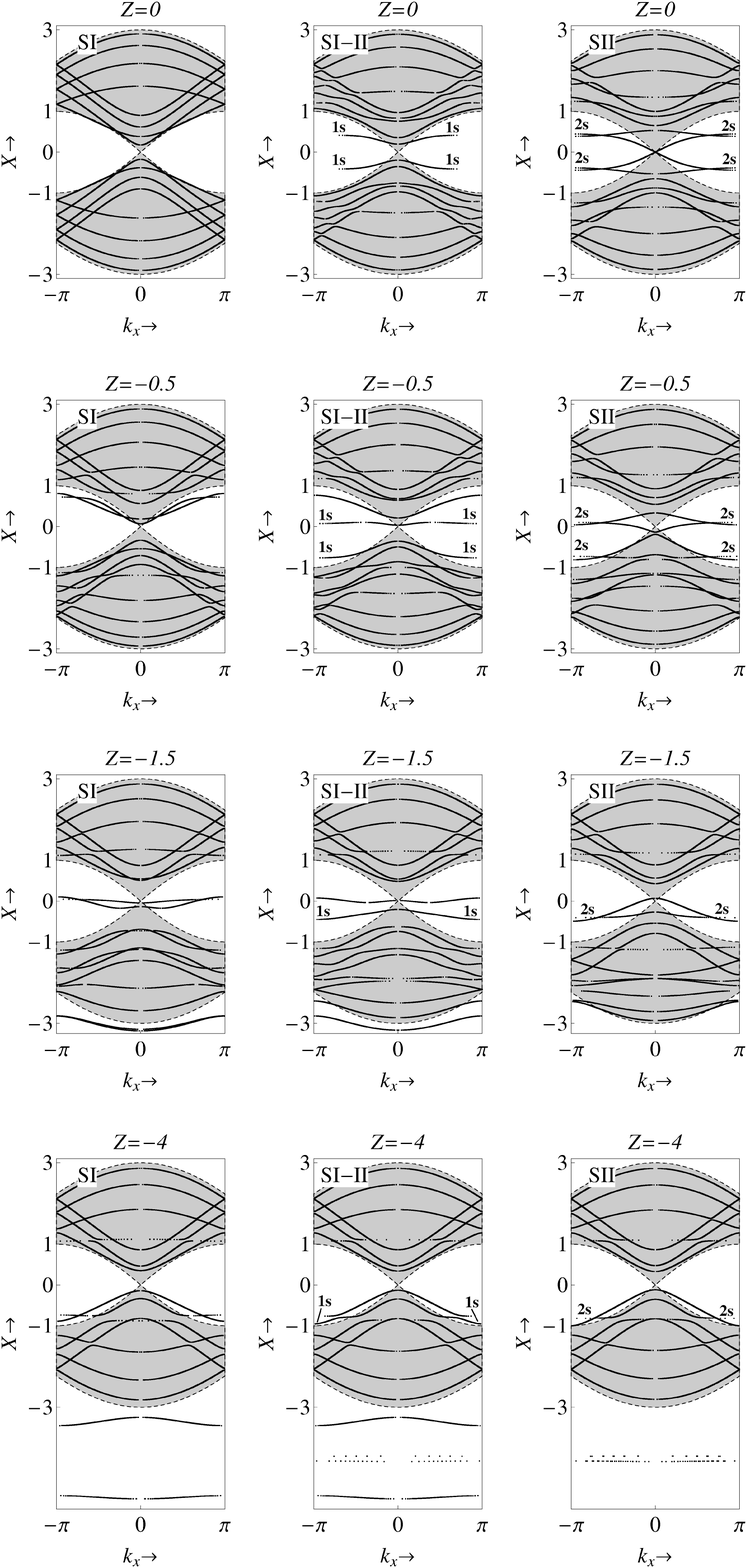}
\caption{\label{fig:f7}Energy spectra of armchair graphene nanoribbons of width $n=5$ and geometries $SI$, $SI-II$ and $SII$ (columns), plotted for increasing absolute value of surface perturbation $Z=0$, $-0.5$, $-1.5$ and $-4$ (rows).}
\end{figure}

As the surface perturbation increases, Shockley states shift towards the gap limit and progressively lose their localization. At the same time, Tamm states emerge from the bands. In the case of strong surface perturbation Tamm states are seen to occur below the lower band, above the upper one, and between the bands, for $\left|k_{y}\right|<\tfrac{2}{3}\pi$  or $\left|k_{y}\right|>\tfrac{2}{3}\pi$  in structures $SI$ and $SII$, respectively. Extreme surface perturbation values cause surface atoms to become isolated. This is equivalent to edge configuration $I$ becoming configuration $II$, and vice versa. This evolution is seen in the dispersion relations obtained for $Z=0$ and for $Z=-3$: with the two branches of strongly localized surface states discarded, the spectrum of structure $SI$ resembles that of structure $SII$. The mechanism of evolution of the $SII$ spectrum into the $SI$ spectrum with growing perturbation is identical. In structure $SI-II$ discarding the surface sites is equivalent to exchanging the edge structures, which leads to the same spectrum.

Figure \ref{fig:f5} shows wave function amplitudes for the first slice of lattice sites in the cell. The assumed values of energy $X$ and wave vector component $k_{y}$ are indicated by dots in surface state branches in Fig. \ref{fig:f4}. The wave function profiles testify the localization of each state at one surface (in the asymmetric structure $SI-II$) or at both surfaces (in the symmetric structures $SI$ and $SII$). States in the symmetric structures $SI$ and $SII$ are easily found to represent even and odd combinations of states localized at a single surface of type $I$ or $II$ in the $SI-II$ structure. The only important features reflecting difference between $S$ and $A$ structures is the symmetry of wave functions. However, the conclusion concerning localization on the surface of type $I$ or $II$ are the same in both cases. In order to prove a localization on both surfaces for structures $AI$ and $AII$ one have to analyse wave functions profiles on a series successive slices. Each of them is localised on one edge only but the direction of localisation changes alternately. 

Figure \ref{fig:f6} presents surface states in nanoribbons of width $n = 50$ (bulk states are hidden for clarity). In a nanoribbon of this width the surfaces are practically isolated from each other, which implies a virtual degeneracy of surface states even near the Dirac points, where the gap is narrow. The surface state spectra of systems of geometries $SI$ and $SII$ correspond to those of a semi-infinite graphene sheet of edge structure $I$ or $II$, respectively. The spectrum of structure $SI-II$ is a superposition of those obtained for $SI$ and $SII$.

\begin{figure}
\centering
\includegraphics[width=3.5in]{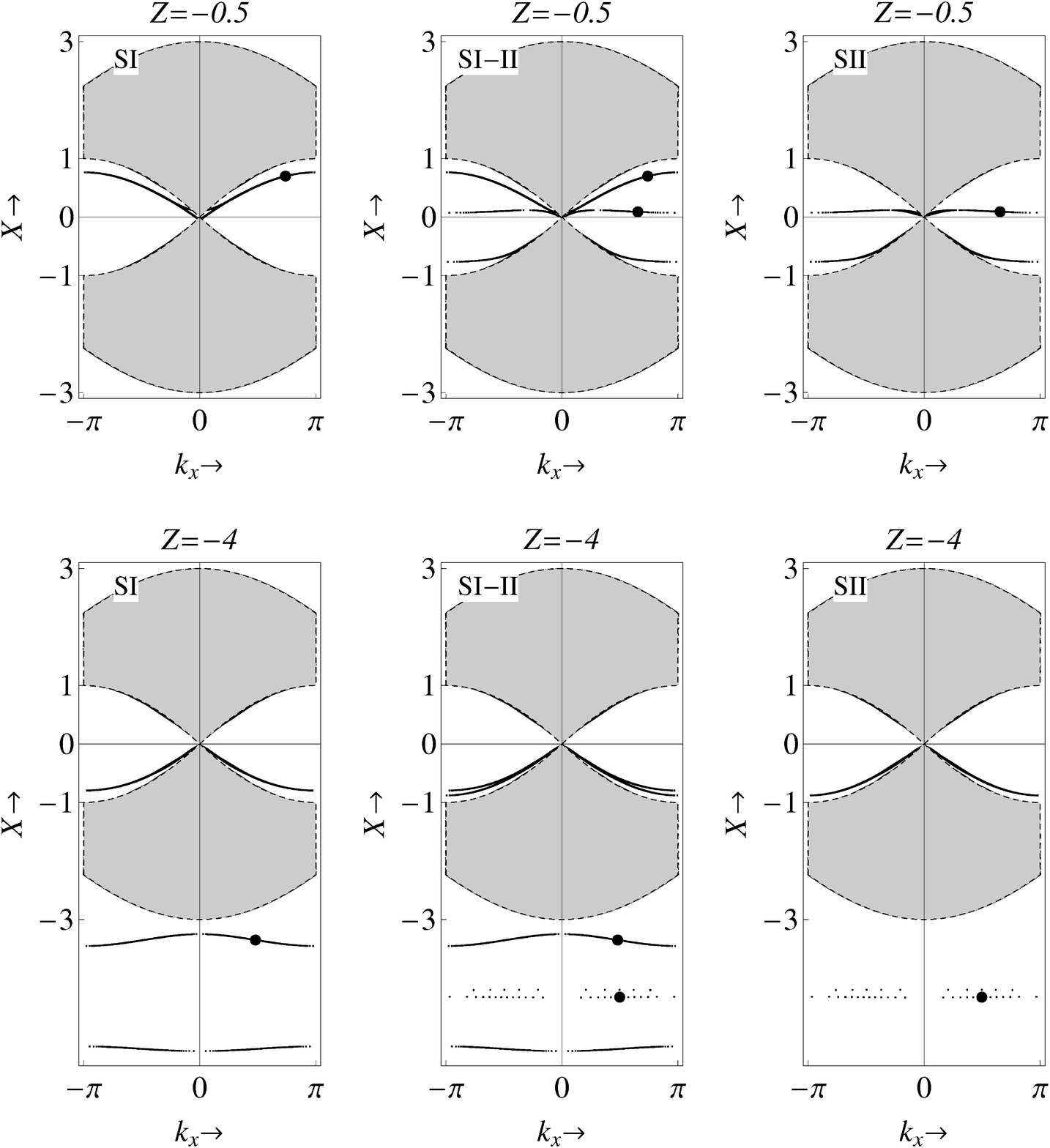}
\caption{\label{fig:f8}Dispersion spectra of surface states in armchair nanoribbons of width $n=15$ and geometries $SI$, $SI-II$ and $SII$ with surface perturbation $Z=-0.5$ and $-4$. Dots in surface state dispersion branches correspond to $k_{y}$ and $X$ values assumed in the wave function plots shown in Fig. \ref{fig:f9}.}
\end{figure}

\begin{figure}
\centering
\includegraphics[width=3.5in]{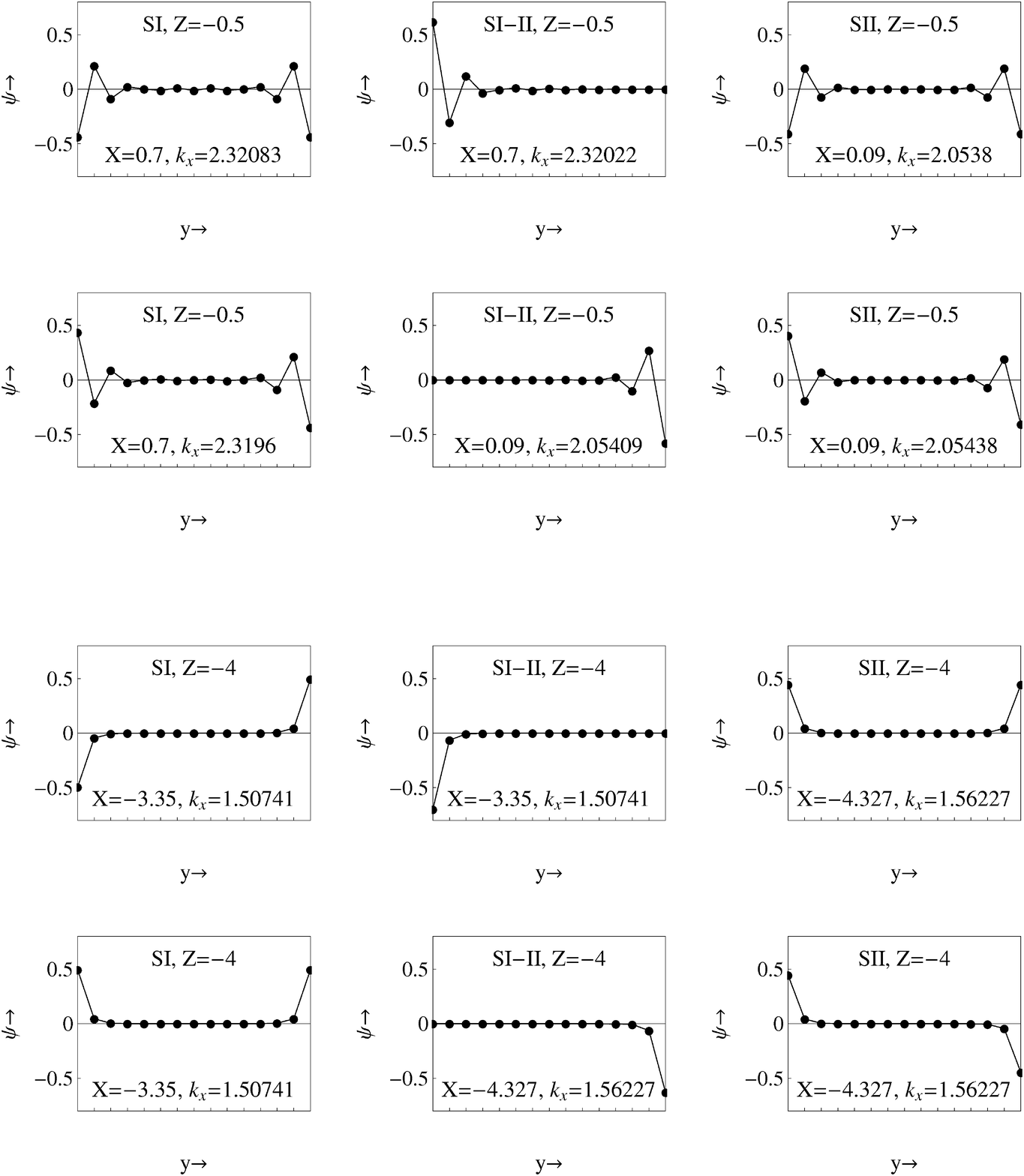}
\caption{\label{fig:f9}Profiles of surface state wave functions plotted across an armchair nanoribbon for the first slice of sites in the nanoribbon cell (cf. Fig. \ref{fig:f2}). The corresponding states are indicated in dispersion spectra plotted in Fig. \ref{fig:f8} by black dots.}
\end{figure}

Results of a similar investigation for armchair nanoribbons are depicted in Fig. \ref{fig:f7}. The presented spectra refer to nanoribbons of edge geometries $SI$, $SI-II$ and $SII$ and width $n = 5$. In each of the studied structures the spectrum comprises 18 dispersion branches. Each surface brings two surface atoms to the nanoribbon cell (gray area in Fig. \ref{fig:f2}), which implies that at most four surface state dispersion branches can occur in the energy gaps. Worthy of notice, and only characteristic of armchair nanoribbons, are energy ranges $1<X<\sqrt{5}$ and $-\sqrt{5}<X<-1$ in which the gap is closed at any value of $k_{x}$. This means no states localized at the nanoribbon surface can occur in these energy ranges.

The spectrum obtained for structure $SI$ indicates that no surface states occur at the type $I$ surface without surface perturbation ($Z = 0$). However, Shockley states are seen to occur at the type $II$ surface (cf. the spectra obtained for structures $SII$ and $SI-II$ at $Z = 0$). Two and four Shockley states occur in structures $SI-II$ and $SII$, respectively, due to the number of type $II$ surfaces in each (one in $SI-II$ and two in $SII$).

The spectra of armchair nanoribbons have the same symmetry with respect to the sign of surface perturbation $Z$ as the spectra obtained for zigzag structures. Only negative values of Z are assumed in this investigation. Increasing the absolute value surface perturbation causes Shockley states to merge into energy bands, with the concurrent induction of Tamm states. When the surface perturbation is strong ($Z=-4$), surface sites in each structure ($SI$, $SI-II$ and $SII$) become isolated, and the system resembles structure $SI$. This is evidenced in the spectrum by the detachment of four surface state branches from the lower band (note in structures $SI$ and $SII$ the detached branches are virtually double-degenerate and overlap).

Figure \ref{fig:f8} shows spectra of surface states in armchair nanoribbons of width $n=15$ (bulk state branches are hidden for clarity). Except for the immediate vicinity of the Dirac point, the surfaces are seen to be well isolated from each other, as evidenced by the close degeneracy of surface states in structures $SI$ and $SII$. By comparing the spectrum of structure $SI-II$ with the spectra of $SI$ and $SII$ it is easy to determine the type of surface at which each surface state localizes in the $SI-II$ system. This becomes even more clear if we compare the wave function profiles depicted in Fig. \ref{fig:f9}. The conclusions as to the parity of the states with respect to the nanoribbon center in the symmetric structures $SI$ and $SII$ and their relation to the wave functions in the asymmetric structure $SI-II$ are the same as in the case of zigzag nanoribbons.

\begin{figure}
\centering
\includegraphics[width=3.5in]{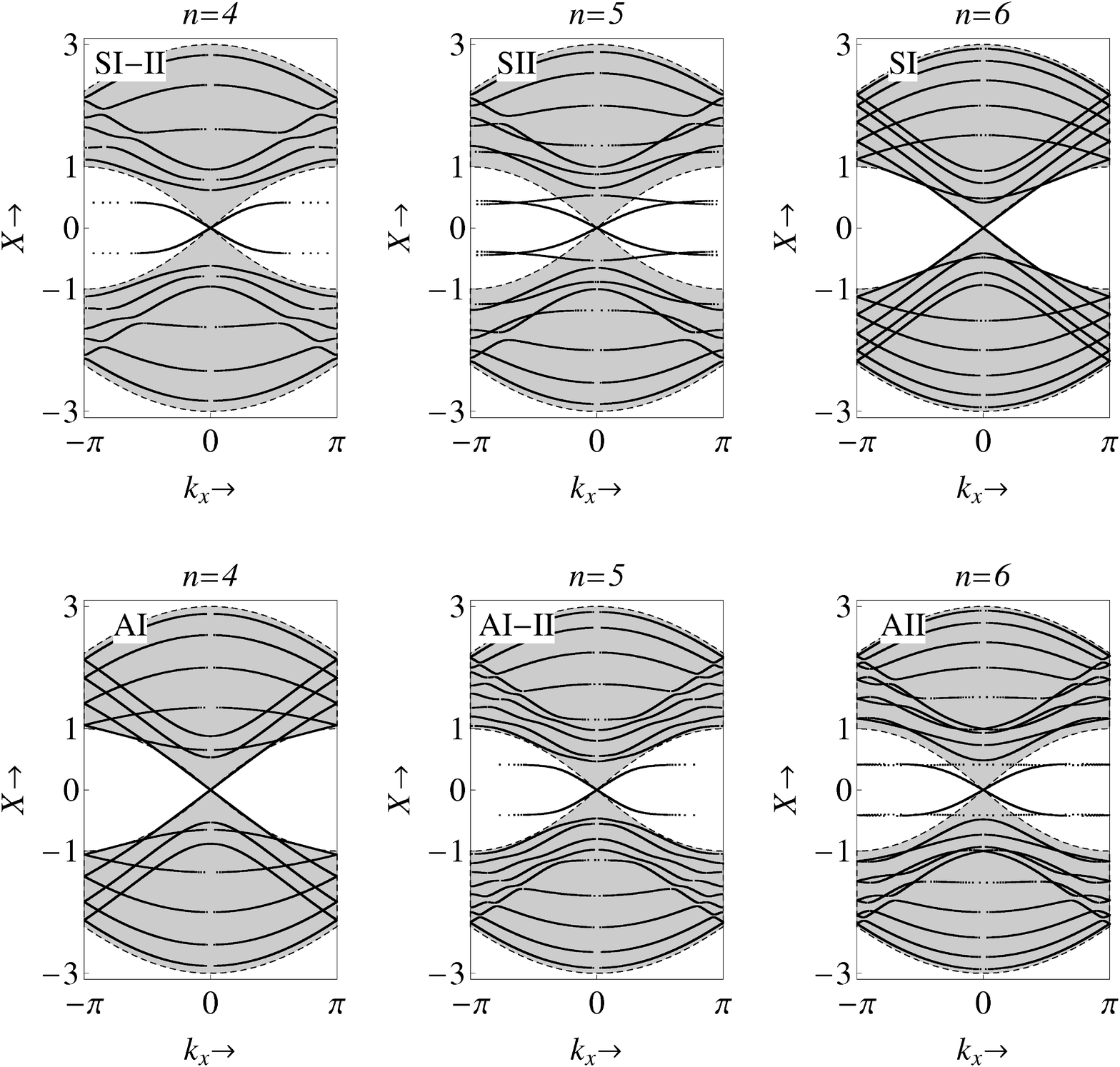}
\caption{\label{fig:f10}Dispersion spectra of armchair nanoribbons with energy gap closed at Dirac point by width adjustment.The surface perturbation is absent ($Z=0$). Note the gap is closed by bulk states in structures $AI$ and $SI$, and by surface states in the other structures.}
\end{figure}

\begin{table}
\caption{\label{tab:t2}Nanoribbon width $n$ corresponding to energy gap closing at Dirac point in armchair nanoribbons of different edge geometry; $m$ is an integer.}
\begin{ruledtabular}
\begin{tabular}{c|cccccc}
structure &AI&AI-II&AII&SI&SI-II&SII \\
\hline
n&3m+1&3m+2&3m&3m&3m+1&3m+2\\
\end{tabular}
\end{ruledtabular}
\end{table}

An interesting effect, which we would like to emphasize, is the energy gap closing around the Dirac point. In AI nanoribbons the gap is closed by bulk states when the nanoribbon has width $n=3m+1$, where $m=1,2,3,\ldots$ Bulk states can also close the energy gap in an $SI$ nanoribbon of width $n=3m$ (which means two slices in the nanoribbon cell comprise $3m$ sites and the other two $3m-1$ sites). In the other structures the gap is closed by surface state branches. The nanoribbon width and the number of atoms per nanoribbon cell for which the energy gap closes near the Dirac point in each of the structures under discussion are specified in Table \ref{tab:t2}. Figure \ref{fig:f10} presents spectra obtained for each structure with nanoribbon width adjusted so that the effect of gap closing can be observed. The dispersion relations plotted in Fig. \ref{fig:f10} correspond to zero surface perturbation.
 
Note the accidental degeneracy due to the intersection of dispersion branches in structures $SI$ and $AI$ (with both edges of type $I$) is partially eliminated in structures $SII$ and $AII$ (with both edges of type $II$) and does not occur at all in structures $SI-II$ and $AI-II$ (with mixed edges). If the nanoribbon is not wide enough, the elimination of the degeneracy due to the intersection of dispersion branches will result in the generation of energy gaps between bulk dispersion branches.

\section{Conclusion}
We have determined the energy spectrum of zigzag and armchair graphene nanoribbons in the tight-binding approximation for   electrons. Two model edge configurations, $I$ and $II$, have been considered, with surface sites having two neighbors or a single neighbor, respectively. A surface perturbation has been allowed for and modeled by a modification of the surface site energy (potential). Shockley states, or surface states which do not require surface perturbation, have been found to occur at edges of both configurations in zigzag nanoribbons. In armchair nanoribbons, only the type $II$ edge configuration permits the occurrence of surface states of this category. The generation of Tamm states by the surface perturbation and its effect on the occurrence of Shockley states have been examined as well. For both structure, a sufficiently strong perturbation destroys Shockley states. Localization of Tamm states is, in general, improved with increasing surface perturbation. The only exceptions are Tamm states in the gap between bands for armchair structure. They disappear in the one of bands for a large surface perturbation.
We have also determined the conditions of energy gap closing at the Dirac point in armchair nanoribbons. In armchair nanoribbons with both edges of type $I$ the gap can be closed by bulk states; otherwise (in structures $II$ or $I-II$) the gap can only be closed by surface state branches.

\begin{acknowledgments}
I. Zozoulenko and A. Shylau are acknowledged for fruitful and helpful discussion. This study was supported by Polish Ministry of Science and Higher Education within the program \it{Support of International Mobility, 2nd edition}.
\end{acknowledgments}


\end{document}